\documentclass[11pt,a4paper]{article}
\usepackage{amssymb}
\usepackage{amsmath}
\usepackage{graphics}
\usepackage{epsfig}
\usepackage{float}
\usepackage{graphicx}
\usepackage{amsfonts}
\usepackage{amsthm}
\usepackage{newlfont}
\usepackage{color}

\begin{document}

\title{$B_{c}$ absorption cross sections by nucleons}
\author{Faisal Akram\thanks{%
faisal.chep@pu.edu.pk (corresponding author)} and M. A. K. Lodhi\thanks{%
a.lodhi@ttu.edu} \\
$^{*}$\textit{Center for High Energy Physics, Punjab University, Lahore,
PAKISTAN} \\
$^{\dag}$\textit{Department of Physics, MS 1051, Texas Tech University,
Lubbock\ TX 79409, USA}}
\maketitle

\begin{abstract}
\noindent The cross sections of $B_{c}$\ absorption by nucleons are
calculated in meson-baryon exchange model using hadronic Lagrangian based on
SU(4)/SU(5) flavor symmetries. The values of different coupling constants
used in the model are obtained from vector meson dominance model, QCD sum
rule or SU(4)/SU(5) flavor symmetries. Calculated values of cross sections
are found to be significantly different from the previous study in which
b-flavored hadron exchange is neglected. These results could be useful in
calculating production rate of $B_{c}$ meson in relativistic heavy ion
collisions.
\end{abstract}


\bigskip


\section{Introduction}

Suppression of $J/\psi $ due to color Debye screening in Quark-Gluon plasma
(QGP) was suggested by T. Matsui and H.\ Satz \cite{Matsui1986}. However,
this suppression may also occur due to interaction of $J/\psi $ with
hadronic comover mainly pions, $\rho $ mesons and nucleons \cite{Gavin1988}.
Due to large density of these comovers the effect of interaction could be
significant even for a relatively small values of absorption cross section,
a few mb \cite{Cassing1997}. Thus, the knowledge of absorption cross
sections is required to interpret the observed suppression of $J/\psi $ in
NA50 experiment at CERN \cite{NA50}. Extensive work has been done to
calculate these cross sections using perturbative QCD \cite{Kharzeev1994},
QCD sum-rule approach \cite{sum-rule}, quark potential models \cite{quark
models} and meson-baryon exchange models based on hadronic Lagrangian \cite%
{Haglin2000,Lin2000,Liu2001}. Bottomonium states are also affected in the
QGP due to color Debye screening \cite{Matsui1986,Vogt1999}. In this case
the related absorption cross sections are calculated using meson-exchange
model in Ref. \cite{Lin2001}. Recently the suppression of ground and excited
states of $\Upsilon $ is observed in Pb+Pb collisions at CMS \cite{cms}. The
observed suppression is expected to be the blend of different effects
including initial and final states interactions with the comovers. Thus we
require the knowledge of the cross sections of these interactions in order
to separate any suppression occurring due to QGP.

In the Refs. \cite{Schro2000,CUP2002,Becattini2005} the studies of
mixed flavor heavy hadrons are also suggested to probe the properties of
QGP. These studies predict an enhancement in the production rate of heavy
hadrons like $B_{c}$ meson, $\Xi _{bc}$, and $\Omega _{ccc}$ baryons due to
QGP. However, once again the knowledge of interaction cross sections are
required to separate any enhancement occurring due to QGP. In this regard $B_{c}$
absorption cross sections by the pions and $\rho $ mesons are recently calculated in
Ref. \cite{Akram2011,Akram2011a}. Calculated cross sections are found to be
in the range 2 to 7 mb and 0.2 to 2 mb for the processes $B_{c}^{+}\pi
\rightarrow DB$ and $B_{c}^{+}\pi \rightarrow D^{\ast }B^{\ast }$
respectively, and 0.6 to 3 mb and 0.05 to 0.3 mb for the processes $%
B_{c}^{+}\rho \rightarrow D^{\ast }B$ and $B_{c}^{+}\rho \rightarrow
DB^{\ast }$ respectively, when the form factor is included. $B_{c}$
absorption cross sections by nucleons have been calculated in Ref. \cite%
{Lodhi2007} using meson-baryon exchange model. These cross sections are
found to have values on the order of few mb. The calculations in Ref. \cite%
{Lodhi2007} included only c-flavored hadron and did not include b-flavored
hadron exchange processes which could significantly change the values of the
cross sections. In this paper, we have calculated these cross sections again
in meson-baryon exchange model and included the effect of b-flavor exchange
as well as anomalous parity interaction.

\noindent This paper is organized as follows. In section II we define
hadronic Lagrangian density and derive the interaction terms relevant for $%
B_{c}$ absorption by nucleons. In section III we produce the amplitudes
of the absorption processes. In section IV we discuss the numerical values
of different coupling constants used in the calculations. In section V we
present results of cross sections and study the effect of uncertainty in
cutoff parameter. In section V the effect of anomalous parity interaction
is discussed. Finally, some concluding remarks are made in the last section.
The paper include an appendix in which derivation of SU(5) invariant
Lagrangian of baryons interaction with mesons is given.

\section{Interaction Lagrangian}

The following processes are studied in this work using meson-baryon exchange
model.%
\begin{equation}
NB_{c}^{+}\rightarrow \Lambda _{c}B,\ \ NB_{c}^{+}\rightarrow \Lambda
_{c}B^{\ast },\ \ NB_{c}^{-}\rightarrow \overline{D}\Lambda _{b},\ \
NB_{c}^{-}\rightarrow \overline{D}^{\ast }\Lambda _{b}  \label{1}
\end{equation}

\noindent It is noted that $B_{c}$ absorption by nucleons processes also
include the channels in which $\Sigma_{c(b)}$ is produced instead of $%
\Lambda_{c(b)}$. The cross sections of these processes may be related to
that of Eq. \ref{1} through isospin symmetry and are not included in this
study. To calculate the cross sections of the processes of Eq. \ref{1}, we
require the following effective interaction Lagrangian densities.
\begin{subequations}
\label{2}
\begin{eqnarray}
\mathcal{L}_{B_{c}BD^{\ast }} &=&ig_{B_{c}BD^{\ast }}D^{\ast \mu
}(B_{c}^{-}\partial _{\mu }B-\partial _{\mu }B_{c}^{-}B)+hc  \label{2a} \\
\mathcal{L}_{B_{c}B^{\ast }D} &=&ig_{B_{c}B^{\ast }D}\overline{B}^{\ast \mu
}(B_{c}^{+}\partial _{\mu }\overline{D}-\partial _{\mu }B_{c}^{+}\overline{D}%
)+hc  \label{2b} \\
\mathcal{L}_{DN\Lambda _{c}} &=&ig_{DN\Lambda _{c}}(\overline{N}\gamma
^{5}\Lambda _{c}\overline{D}+D\overline{\Lambda }_{c}\gamma ^{5}N)
\label{2c} \\
\mathcal{L}_{D^{\ast }N\Lambda _{c}} &=&g_{D^{\ast }N\Lambda _{c}}(\overline{%
N}\gamma ^{\mu }\Lambda _{c}\overline{D}^{\ast }+D^{\ast }\overline{\Lambda }%
_{c}\gamma ^{\mu }N)  \label{2d} \\
\mathcal{L}_{BN\Lambda _{b}} &=&ig_{BN\Lambda _{b}}(\overline{N}\gamma
^{5}\Lambda _{b}B+\overline{B}\overline{\Lambda }_{b}\gamma ^{5}N)
\label{2e} \\
\mathcal{L}_{B^{\ast }N\Lambda _{b}} &=&g_{B^{\ast }N\Lambda _{b}}(\overline{%
N}\gamma _{\mu }\Lambda _{b}B^{\ast \mu }+\overline{B}^{\ast \mu }\overline{%
\Lambda }_{b}\gamma _{\mu }N)  \label{2f} \\
\mathcal{L}_{B_{C}\Lambda _{c}\Lambda _{b}} &=&ig_{B_{C}\Lambda _{c}\Lambda
_{b}}(\overline{\Lambda }_{c}\gamma ^{5}\Lambda _{b}B_{c}^{+}+\overline{%
\Lambda }_{b}\gamma ^{5}\Lambda _{c}B_{c}^{-})  \label{2g}
\end{eqnarray}

\noindent Where,

\end{subequations}
\begin{eqnarray}
D &=&\left(
\begin{array}{cc}
D^{0} & D^{+}%
\end{array}%
\right) ,\overline{D}=\left(
\begin{array}{cc}
\overline{D}^{0} & D^{-}%
\end{array}%
\right) ^{T},D_{\mu }^{\ast }=\left(
\begin{array}{cc}
D_{\mu }^{\ast 0} & D_{\mu }^{\ast +}%
\end{array}%
\right) ,  \notag \\
B &=&\left(
\begin{array}{cc}
B^{+} & B^{0}%
\end{array}%
\right) ^{T},B_{\mu }^{\ast }=\left(
\begin{array}{cc}
B_{\mu }^{\ast +} & B_{\mu }^{\ast 0}%
\end{array}%
\right) ^{T},  \notag \\
N &=&\left(
\begin{array}{c}
p \\
n%
\end{array}%
\right)
\end{eqnarray}

\noindent Pseudoscalar-pseudoscalar-vector meson (PPV) couplings given in
Eqs. \ref{2a} and \ref{2b} are obtained from the hadronic Lagrangian based
on SU(5) gauge symmetry introduced in Ref. \cite{Lin2001}. In this Lagrangian the coupling constant of different PPV, VVV and PPVV couplings are expressed in terms of single
coupling constant $g$. For example, the coupling constants $g_{\pi DD^{\ast
}},\ g_{\pi BB^{\ast }},$ $g_{B_{c}BD^{\ast }},$ and $g_{B_{c}B^{\ast }D}$
are given as following. 
\begin{equation}
g_{\pi DD^{\ast }}=g_{\pi BB^{\ast }}=\frac{g}{4},\text{ }g_{B_{c}BD^{\ast
}}=g_{B_{c}B^{\ast }D}=\frac{g}{2\sqrt{2}}  \label{3}
\end{equation}

\noindent All the mass terms of vector mesons, which break the SU(5)
symmetry, are added directly in the Lagrangian density. Thus, it is expected
that SU(5) symmetry relations given in Eq. \ref{3} are violated.

\noindent Baryon-baryon-pseudoscalar meson (BBP) and baryon-baryon-vector
meson (BBV) couplings given in Eqs. \ref{2c} to \ref{2g} can be obtained
from the following SU(5) invariant Lagragians \cite{Liu2001,Okubo1975}.%
\begin{eqnarray}
\mathcal{L}_{PBB} &=&g_{P}(a\phi ^{\ast \alpha \mu \nu }\gamma ^{5}P_{\alpha
}^{\beta }\phi _{\beta \mu \nu }+b\phi ^{\ast \alpha \mu \nu }\gamma
^{5}P_{\alpha }^{\beta }\phi _{\beta \nu \mu })  \label{4} \\
\mathcal{L}_{VBB} &=&ig_{V}(c\phi ^{\ast \alpha \mu \nu }\gamma V_{\alpha
}^{\beta }\phi _{\beta \mu \nu }+d\phi ^{\ast \alpha \mu \nu }\gamma
V_{\alpha }^{\beta }\phi _{\beta \nu \mu })  \label{5}
\end{eqnarray}

\noindent where all the indices run from 1 to 5. The tensors $P_{\alpha
}^{\beta }$ and $V_{\alpha }^{\beta }$ are defined by the pseduscalar and
vector meson matrices given in Ref. \cite{Lin2001} and the tensor $\phi
^{\alpha \mu \nu } $ defines the $J^{P}=\frac{1}{2}^{+}$ baryons which
belongs to 40-plet states in SU(5) quark model (see the Appendix). The
Lagrangian density of Eq. \ref{4} defines all the BBP couplings in terms of
universal coupling $g_{P}$ and the constants $a$ and $b$. Similarly the
Lagrangian density of Eq. \ref{5} defines all the BBV couplings in terms of
universal coupling $g_{V}$ and the constants $c$ and $d$. For the coupling
constants $g_{\pi NN}\,,~g_{\rho NN},~g_{KN\Lambda },~g_{K^{\ast }N\Lambda }$
and given in the Eqs. \ref{2c} to \ref{2g}, we obtain the following results.

\begin{eqnarray}
g_{\pi NN} &=&\frac{1}{\sqrt{2}}g_{P}(a-\frac{5}{4}b),\text{ \ }%
g_{B_{c}\Lambda _{c}\Lambda _{b}}=\frac{3}{4}g_{P}(a-b),  \notag \\
g_{KN\Lambda } &=&g_{DN\Lambda _{c}}=g_{BN\Lambda _{b}}=\frac{3\sqrt{6}}{8}%
g_{P}(b-a)  \label{6} \\
g_{\rho NN} &=&\frac{1}{\sqrt{2}}g_{V}(c-\frac{5}{4}d),~\ g_{K^{\ast
}N\Lambda }=g_{D^{\ast }N\Lambda _{c}}=g_{B^{\ast }N\Lambda _{b}}=\frac{3%
\sqrt{6}}{8}g_{V}(d-c)  \label{7}
\end{eqnarray}

\noindent SU(5) flavor symmetry in badly broken due to large variation in
the related quark masses. Thus, these symmetry relations are also expected
to be violated. It is noted that SU(4) flavor symmetry also produces the
same relations as given in Eqs. \ref{6} \& \ref{7} for couplings of the
hadrons containing $u,~d,~s$ and $c$ quarks \cite{Liu2001}.

\section{Amplitudes of $B_{c}$ absorption processes}

\noindent Shown in Fig. 1 are the Feynman diagrams of the four processes
given by Eq. \ref{1}. Corresponding to each process, we have two diagrams.
In all $t$ and $u$ channel diagrams $c$ and $b$ flavors are exchanged
respectively.

\begin{figure}[!h]
\begin{center}
\includegraphics[angle=0,width=0.80\textwidth]{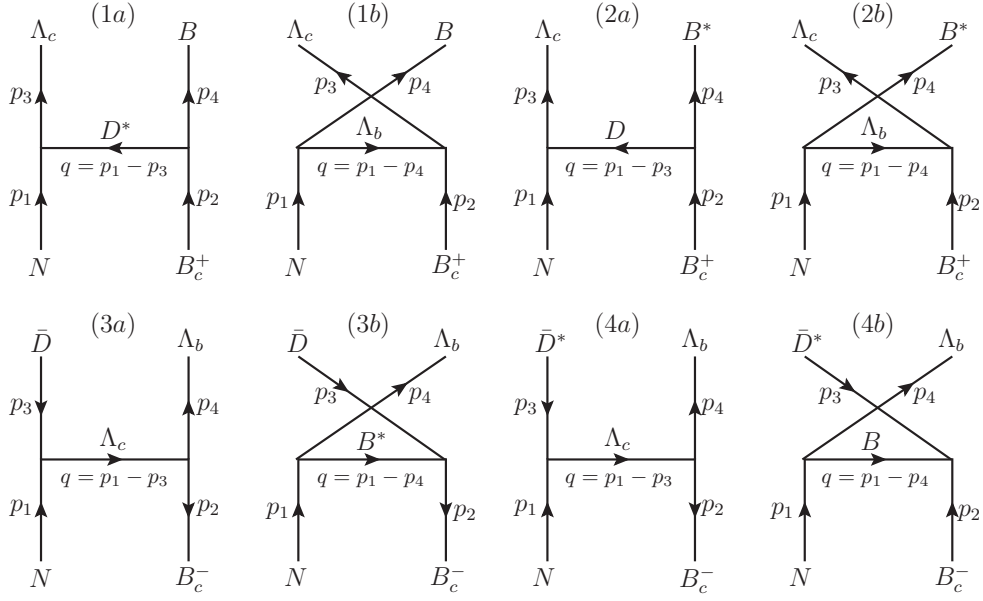}
\end{center}
\caption{Feynman diagram for $B_{c}$ absorption processes:$%
~(1)~NB_{c}^{+}\rightarrow \Lambda _{c}B,\ (2)\ NB_{c}^{+}\rightarrow
\Lambda _{c}B^{\ast },\ (3)\ NB_{c}^{-}\rightarrow \overline{D}\Lambda
_{b},\ (4)\ NB_{c}^{-}\rightarrow \overline{D}^{\ast }\Lambda _{b}$}
\label{fig1}
\end{figure}

\noindent Scattering amplitudes of these diagrams are given by,%
\begin{eqnarray}
M_{1a} &=&-g_{D^{\ast }N\Lambda _{c}}g_{B_{c}BD^{\ast }}(-p_{4}-p_{2})_{\mu }%
\frac{-i}{t-m_{D^{\ast }}^{2}}\left( g^{\mu \nu }-\frac{(p_{1}-p_{3})^{\mu
}(p_{1}-p_{3})^{\nu }}{m_{D^{\ast }}^{2}}\right)  \notag \\
&&\times ~\overline{u}_{\Lambda _{c}}(p_{3})\gamma _{\nu }u_{N}(p_{1})
\notag \\
M_{1b} &=&g_{BN\Lambda _{b}}g_{B_{c}\Lambda _{c}\Lambda _{b}}\overline{u}%
_{\Lambda _{c}}(p_{3})\gamma ^{5}\left( i\frac{(p_{1}-p_{4}).\gamma
+m_{\Lambda _{b}}}{u-m_{\Lambda _{b}}^{2}}\right) \gamma ^{5}u_{N}(p_{1})
\label{8}
\end{eqnarray}

\bigskip
\begin{eqnarray}
M_{2a} &=&ig_{DN\Lambda _{c}}g_{B_{c}B^{\ast }D}(p_{4}-2p_{2})_{\mu }\frac{i%
}{t-m_{D}^{2}}\overline{u}_{\Lambda _{c}}(p_{3})\gamma
^{5}u_{N}(p_{1})\varepsilon _{B^{\ast }}^{\mu }(p_{4})  \notag \\
M_{2b} &=&-ig_{B^{\ast }N\Lambda _{b}}g_{B_{c}\Lambda _{c}\Lambda _{b}}%
\overline{u}_{\Lambda _{c}}(p_{3})\gamma ^{5}\left( i\frac{%
(p_{1}-p_{4}).\gamma +m_{\Lambda _{b}}}{u-m_{\Lambda _{b}}^{2}}\right)
\gamma _{\mu }u_{N}(p_{1})\varepsilon _{B^{\ast }}^{\mu }(p_{4})  \label{9}
\end{eqnarray}

\bigskip
\begin{eqnarray}
M_{3a} &=&g_{DN\Lambda _{c}}g_{B_{c}\Lambda _{c}\Lambda _{b}}\overline{u}%
_{\Lambda _{b}}(p_{4})\gamma ^{5}\left( i\frac{(p_{1}-p_{3}).\gamma
+m_{\Lambda _{c}}}{t-m_{\Lambda _{c}}^{2}}\right) \gamma ^{5}u_{N}(p_{1})
\notag \\
M_{3b} &=&-g_{B^{\ast }N\Lambda _{b}}g_{B_{c}B^{\ast }D}(-p_{3}-p_{2})_{\mu }%
\frac{-i}{u-m_{B^{\ast }}^{2}}\left( g^{\mu \nu }-\frac{(p_{1}-p_{4})^{\mu
}(p_{1}-p_{4})^{\nu }}{m_{B^{\ast }}^{2}}\right)  \notag \\
&&\times ~\overline{u}_{\Lambda _{b}}(p_{4})\gamma _{\nu }u_{N}(p_{1})
\label{10}
\end{eqnarray}

\bigskip

\begin{eqnarray}
M_{4a} &=&-ig_{D^{\ast }N\Lambda _{c}}g_{B_{c}\Lambda _{c}\Lambda _{b}}%
\overline{u}_{\Lambda _{b}}(p_{4})\gamma ^{5}\left( i\frac{%
(p_{1}-p_{3}).\gamma +m_{\Lambda _{c}}}{t-m_{\Lambda _{c}}^{2}}\right)
\gamma _{\mu }u_{N}(p_{1})\varepsilon _{D^{\ast }}^{\mu }(p_{3})  \notag \\
M_{4b} &=&ig_{BN\Lambda _{b}}g_{B_{c}BD^{\ast }}(p_{3}-2p_{2})_{\mu }\frac{i%
}{u-m_{B}^{2}}\overline{u}_{\Lambda _{b}}(p_{4})\gamma
^{5}u_{N}(p_{1})\varepsilon _{D^{\ast }}^{\mu }(p_{3})  \label{11}
\end{eqnarray}

\noindent Total amplitude of each process is given by,%
\begin{equation}
M_{i}=M_{ia}+M_{ib},~\ \ \ \ \ \ \forall ,~i=1,2,3,4  \label{12}
\end{equation}

\noindent Using the total amplitudes given in Eq. \ref{12}, we calculate
unpolarized and isospin averaged cross sections. The required isospin factor
in this case is simply 1 for all four processes. It is noted that in the
study of the processes of the Eq. \ref{1}, we do not include any diagram in
which $\Sigma_{c}$ or $\Sigma_{b}$ particle is exchanged. These diagram
require $B_{c}\Sigma_{b}\Lambda_{c}$ and $B_{c}\Sigma_{c}\Lambda_{b}$
couplings in addition to $DN\Sigma_{c}$, $BN\Sigma_{b}$, $D^{*}N\Sigma_{c}$
and $B^{*}N\Sigma_{b}$ couplings. These couplings of $B_{c}$ meson violate
isospin (I) and also not produced by the SU(5) invariant Lagrangian given in
Eq. \ref{4}. Therefore, It is a good approximation to neglect $\Sigma_{c}$
or $\Sigma_{b}$ exchange diagrams for the processes given in Eq. \ref{1}.

\section{Numerical values of coupling constants}

\noindent The values of the couplings $g_{B_{c}BD^{\ast }}$ and $%
g_{B_{c}B^{\ast }D}$ are fixed by using $g_{\Upsilon BB}=13.3$, which is
obtained using vector meson dominance (VMD) model in ref. \cite{Lin2001} and
SU(5) symmetry result $g_{B_{c}BD^{\ast }}=g_{B_{c}B^{\ast }D}=\frac{2}{%
\sqrt{5}}g_{\Upsilon BB}$ \cite{Lodhi2007}. In this way we obtain $%
g_{B_{c}BD^{\ast }}=g_{B_{c}B^{\ast }D}=11.9$. The couplings $g_{DN\Lambda
_{c}}$ and $g_{D^{\ast }N\Lambda _{c}} $ can be fixed by using SU(5)/SU(4)
symmetry relations $g_{KN\Lambda }=g_{DN\Lambda _{c}}$ and $g_{K^{\ast
}N\Lambda }=g_{D^{\ast }N\Lambda _{c}}$ given in Eqs. \ref{6} \& \ref{7} and
the empirical values of the couplings $g_{KN\Lambda }$ and $g_{K^{\ast
}N\Lambda }$ given in ref. \cite{Stoks1999}. In this way we obtain the
following results,%
\begin{equation}
g_{DN\Lambda _{c}}=13.1,\ \ \ \ \ \ \ \ g_{D^{\ast }N\Lambda _{c}}=4.3
\label{13}
\end{equation}

\noindent Whereas, the QCD sum-rule approach gives the following values of
these couplings \cite{Duraes2001}.%
\begin{equation}
\left\vert g_{DN\Lambda _{c}}\right\vert =7.9,\ \ \ \ \ \ \ \ \left\vert
g_{D^{\ast }N\Lambda _{c}}\right\vert =7.5  \label{14}
\end{equation}

\noindent Due to significant difference between the values given in Eqs. \ref%
{13} \& \ref{14}, we use both of them separately to study their effect on
the calculated cross sections. However, It is noted that the values given in
Eq. \ref{13} are less reliable due to the effect of breaking of SU(5)/SU(4)
flavor symmetries. There are no empirically fitted values available for the
couplings $g_{B_{c}\Lambda _{c}\Lambda _{b}},~g_{BN\Lambda _{b}}$ and $%
g_{B^{\ast }N\Lambda _{b}}$, thus we use SU(5) symmetry relations given in
Eqs. \ref{6} \& \ref{7}, which implies,%
\begin{equation}
g_{B_{c}\Lambda _{c}\Lambda _{b}}=-\frac{2}{\sqrt{6}}g_{DN\Lambda _{c}},~\ \
g_{BN\Lambda _{b}}=g_{DN\Lambda _{c}},\ \ \ g_{B^{\ast }N\Lambda
_{b}}=g_{D^{\ast }N\Lambda _{c}}\ \ \   \label{15}
\end{equation}

\noindent The values of $g_{DN\Lambda _{c}}$ \& $g_{D^{\ast }N\Lambda _{c}}$
in Eq. \ref{13} give $g_{B_{c}\Lambda _{c}\Lambda _{b}}=-10.7,\,\
g_{BN\Lambda _{b}}=13.1$ \& $g_{B^{\ast }N\Lambda _{b}}=4.3$, whereas the
values in Eq. \ref{14} give $g_{B_{c}\Lambda _{c}\Lambda _{b}}=-6.5,\,\
g_{BN\Lambda _{b}}=7.9$ \& $g_{B^{\ast }N\Lambda _{b}}=7.5$. Where, we
choose the sign of the couplings $g_{DN\Lambda _{c}}$ \& $g_{D^{\ast
}N\Lambda _{c}}$ in accordance with the Eq. \ref{13}. Two sets of the values
of the coupling constants used in this paper and methods of obtaining them
are summarized in Table 1.

\begin{table}[tbp] \centering%

\begin{tabular}{c|c|c|c|c}
\hline\hline
& \multicolumn{2}{c}{Set 1}  \vline & \multicolumn{2}{c}{Set 2}  \\ \hline
Coupling Constant & Value & Method of derivation & Value & Method of derivation \\ \hline
$g_{B_{c}BD^{\ast }}$ \& $g_{B_{c}B^{\ast }D}$ & 11.9 & VMD, SU(5) & 11.9 &
VMD, SU(5) \\
$g_{DN\Lambda _{c}}$ & 13.1 & $g_{KN\Lambda }$, SU(4)/SU(5) & 7.9 & QCD sum-rule
\\
$g_{D^{\ast }N\Lambda _{c}}$ & 4.3 & $g_{K^{\ast }N\Lambda }$, SU(4)/SU(5) & 7.5 &
QCD sum-rule \\
$g_{BN\Lambda _{b}}$ & 13.1 & $g_{DN\Lambda _{c}}$, SU(5) & 7.9 & $%
g_{DN\Lambda _{c}}$, SU(5) \\
$g_{B^{\ast }N\Lambda _{b}}$ & 4.3 & $g_{D^{\ast }N\Lambda _{c}}$, SU(5) &
7.5 & $g_{D^{\ast }N\Lambda _{c}}$, SU(5) \\
$g_{B_{c}\Lambda _{c}\Lambda _{b}}$ & -10.7 & $g_{DN\Lambda _{c}}$, SU(5) &
-6.5 & $g_{DN\Lambda _{c}}$, SU(5) \\ \hline\hline
\end{tabular}%
\caption{Two sets of the values of coupling constants used in this paper.}%
\label{table1}%
\end{table}%

\section{Results and Discussion}

\begin{figure}[!h]
\begin{center}
\includegraphics[angle=0,width=0.45\textwidth]{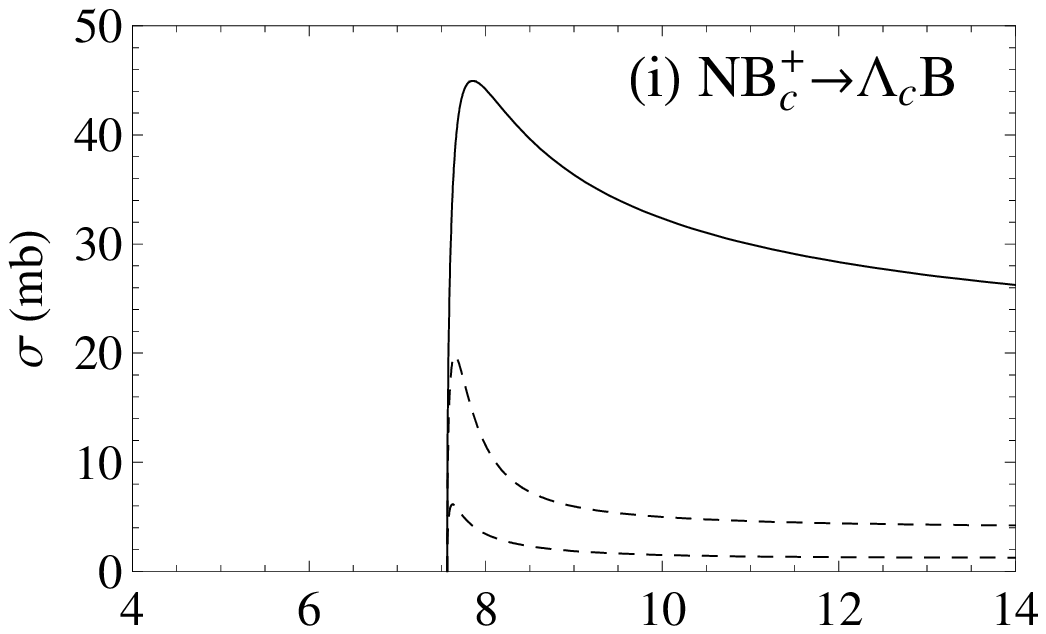} \label{fig2a} %
\includegraphics[angle=0,width=0.45\textwidth]{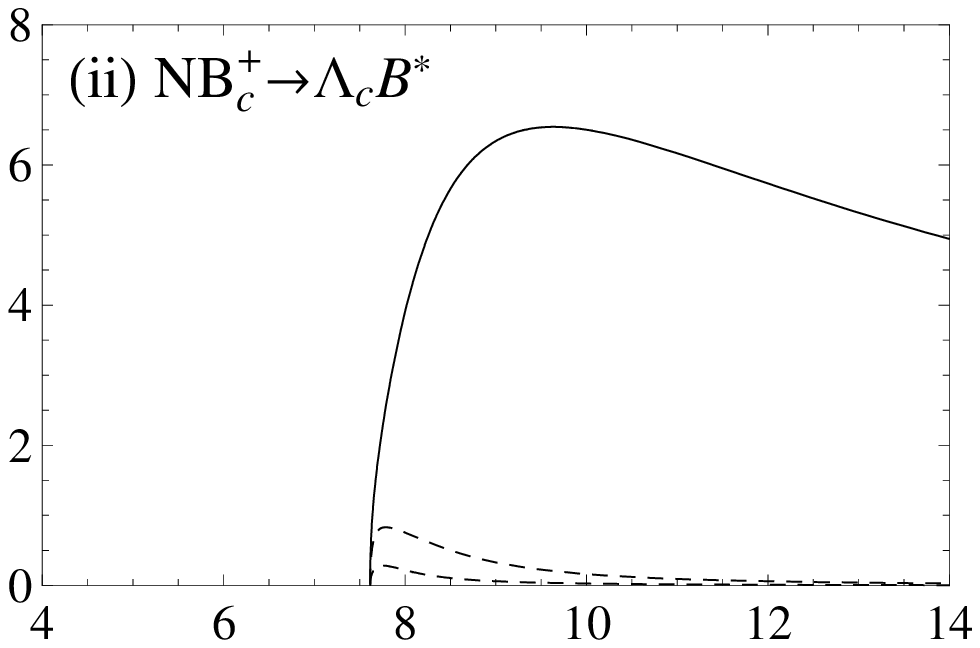} \label{fig2b}
\includegraphics[angle=0,width=0.45\textwidth]{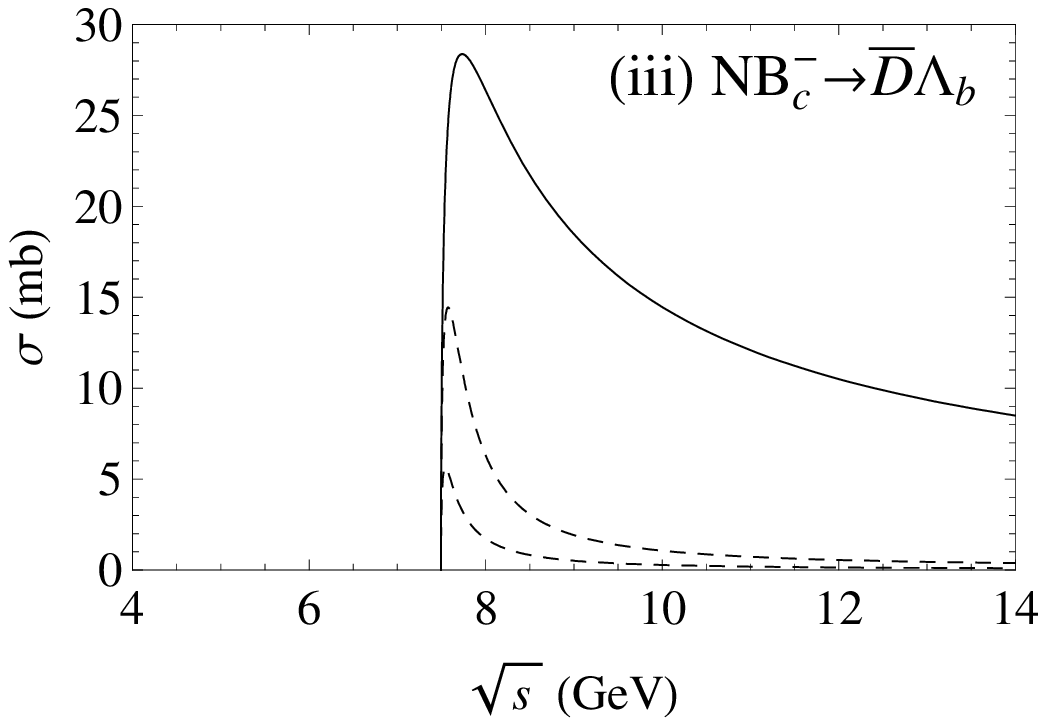} \label{fig2c} %
\includegraphics[angle=0,width=0.45\textwidth]{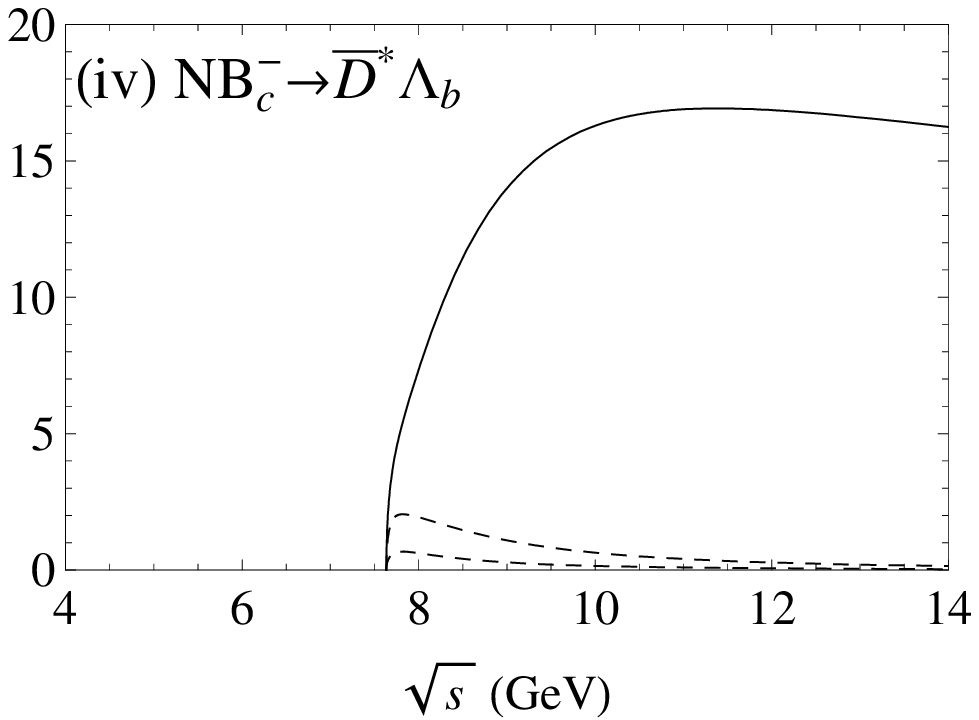} \label{fig2d}

\end{center}
\caption{$B_{c}$ absorption cross sections of the four processes using the values of the couplings given in set 1. Solid and dashed
curves represent cross sections without and with form factor respectively. Lower and upper dashed curves are with cutoff parameter
$\Lambda=1$ and $\Lambda=2$ GeV respectively.}
\label{fig2}
\end{figure}

\noindent Shown in Fig. 2 are the $B_{c}$ absorption cross sections by
nucleons for the four processes given in Eq. \ref{1}, as function of total
center of mass (c.m) energy. These cross sections are obtained using the
values of couplings given in set 1. Solid and dashed curves in these figures
represent cross sections without and with form factors. Form factors are
required to include the effect of finite size of interacting hadrons. In the
present work we use the following monopole form factor.
\begin{equation}
f_{3}=\frac{\Lambda ^{2}}{\Lambda ^{2}+\overline{q}^{2}},
\end{equation}
\noindent where $\Lambda $ is cutoff parameter and $\overline{q}^{2}$ is
squared three momentum transfer in c.m frame. In Ref. \cite{Akram2011}, it
is found that for charm and bottom mesons $\Lambda$ could be in the range $%
1.2-1.8$ GeV. However, we adopt a more conservative view and vary it from 1
to 2 GeV to study the uncertainty in the cross sections due to cutoff
parameter.

\begin{figure}[!h]
\begin{center}
\includegraphics[angle=0,width=0.45\textwidth]{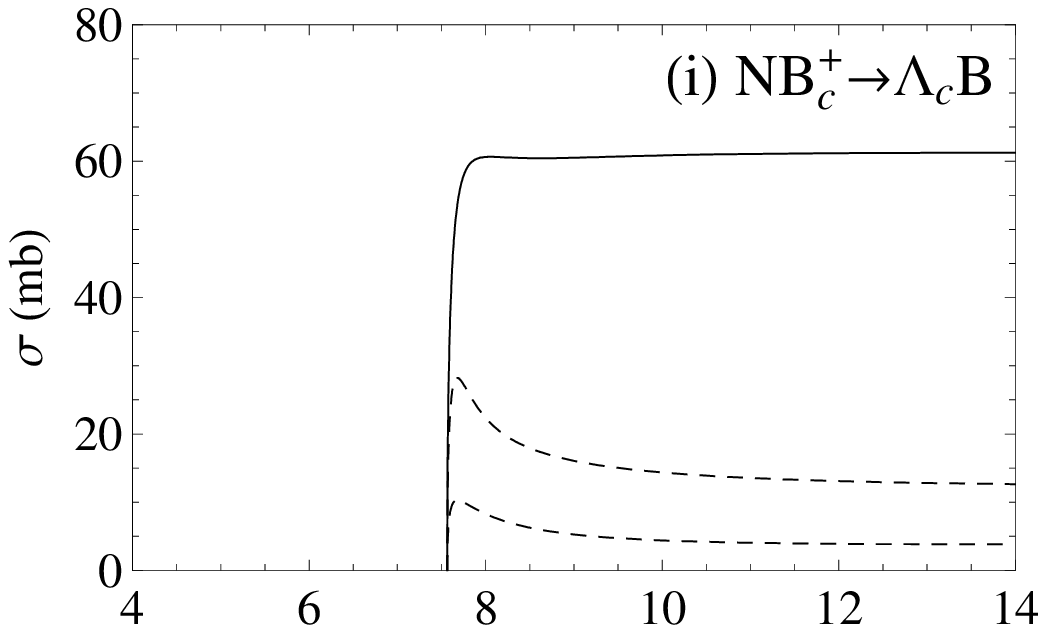} \label{fig3a} %
\includegraphics[angle=0,width=0.45\textwidth]{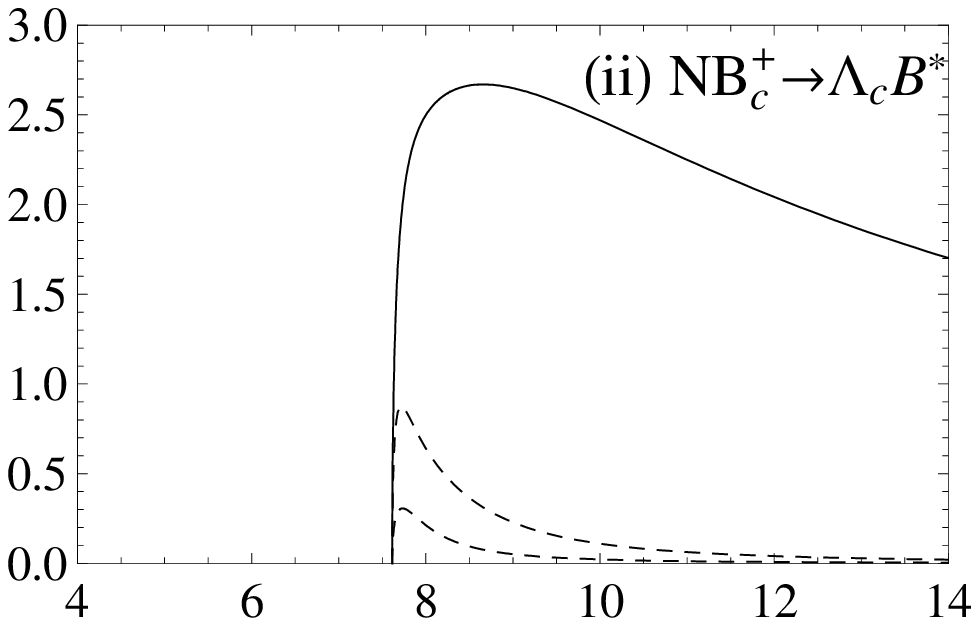} \label{fig3b}
\includegraphics[angle=0,width=0.45\textwidth]{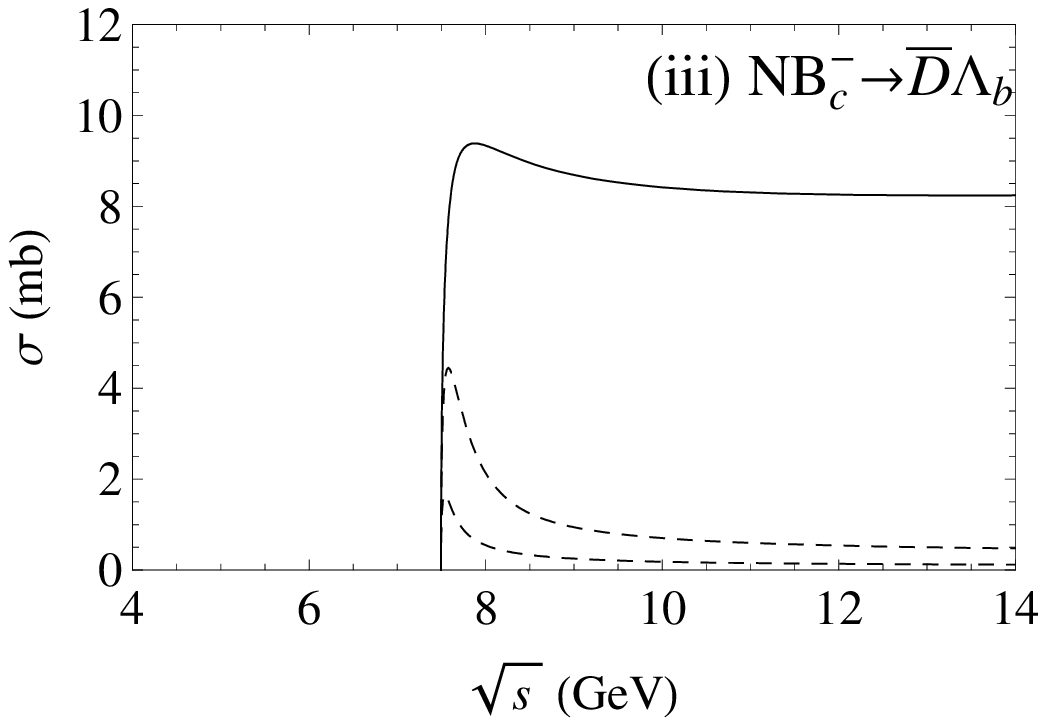} \label{fig3c} %
\includegraphics[angle=0,width=0.45\textwidth]{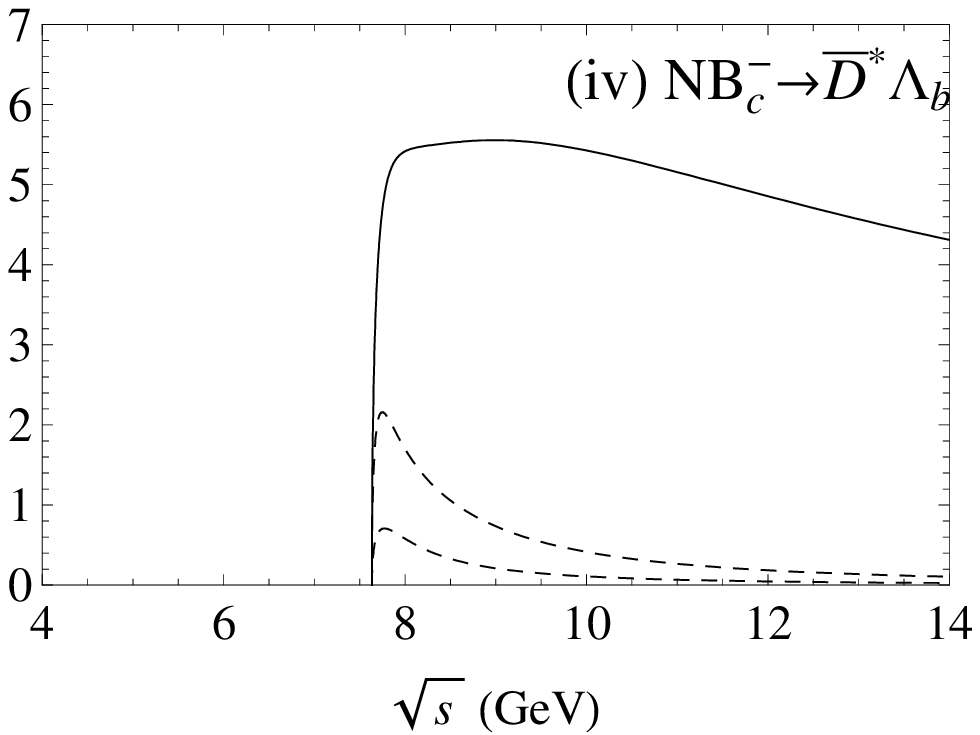} \label{fig3d}

\end{center}
\caption{$B_{c}$ absorption cross sections of the four processes using the values of the couplings given in set 2. Solid and dashed
curves represent cross sections without and with form factor respectively. Lower and upper dashed curves are with cutoff parameter
$\Lambda=1$ and $\Lambda=2$ GeV respectively.}

\label{fig3}
\end{figure}

Fig. 2 shows that, for the processes $(i)$ $NB_{c}^{+}\rightarrow \Lambda
_{c}B,(ii)\ NB_{c}^{+}\rightarrow \Lambda _{c}B^{\ast },(iii)\
NB_{c}^{-}\rightarrow \overline{D}\Lambda _{b},(iv)\ NB_{c}^{-}\rightarrow
\overline{D}^{\ast }\Lambda _{b}$ the cross sections roughly vary $2-5$ mb, $%
0.05-0.3$ mb, $0.1-2$ mb, and $0.1-1$ mb respectively in the most part of
the energy scale \footnote{%
These approximate variations are defined for $\sqrt{s}\geq9$ GeV}, when the
cutoff parameter $\Lambda $ is between $1-2$ GeV. Relatively high
suppression due to cutoff in the processes 2 and 4 is due to higher values
of the masses of vector mesons $D^{\ast }$ and $B^{\ast }$. In Fig. 3, we
present the plots of cross sections using the values of couplings given in
the set 2. Again the solid and dashed curves in these figures represent
cross sections without and with form factors respectively. These cross
sections roughly vary $5-15$ mb, $0.05-0.2$ mb, $0.1-1$ mb and $0.1-0.6$ mb
in the most part of energy scale, for the processes $(i)$ to $(iv)$, when
the cutoff $\Lambda $ is between $1-2$ GeV. In table 2, we present the
comparison of peak values of the cross sections for two sets of the coupling
values. These results show that the values of set 2 increase the peak values
of the cross sections of the process 1 by the factor of $\sim 1.5$ and
decrease of the process 3 by the factor $\sim 4$, whereas the peak values of
the processes 2 and 4 almost remain unchanged. In order to study the effect
of b-flavored hadron exchange, we also present in the table 2, the
comparison of the peak values with and without b-flavor exchange diagrams.
The results show that the b-flavor exchange between interacting hadrons
significantly increases the peak values of the cross sections of the first
three processes for the coupling set 1 and processes (ii) and (iii) for the
set 2. Generally, the contribution of a diagram at tree level depend upon
the mass of the exchange particle, coupling product and the form of
amplitude. Higher value of the mass of the exchange particle tends to
decrease the contribution of a diagram. However, in some case the other
factors like the higher value of the coupling product or the form of
amplitude may significantly increase the contribution even when mass of the
exchange particle is increased. In our case two contributing amplitudes for
each process have different form and contain different coupling product.
Thus, the mere fact that the mass of bottom-hadron is higher than
charm-hadron does not imply that contribution of the bottom-exchange
diagrams is lesser than charm-exchange diagrams.

\begin{table}[tbp] \centering%

\begin{tabular}{c|c|c|c|c|c}
\hline\hline
& & \multicolumn{2}{c}{Set 1}  \vline & \multicolumn{2}{c}{Set 2} \\ \cline{3-6}
& & $\Lambda=1$ GeV & $\Lambda=2$ GeV & $\Lambda=1$ GeV & $\Lambda=2$ GeV \\ \hline
& with b-exchange & 6 mb & 19 mb & 10 mb & 28 mb \\[-1ex]
\raisebox{1.5ex}{$NB_{c}^{+}\rightarrow \Lambda _{c}B$} & without b-exchange & 3 mb & 7 mb & 9 mb & 23 mb \\ \hline
& with b-exchange & 0.25 mb & 0.8 mb & 0.3 mb & 0.75 mb \\[-1ex]
\raisebox{1.5ex}{$NB_{c}^{+}\rightarrow \Lambda _{c}B^{*}$} & without b-exchange & 0.08 mb & 0.6 mb & 0.03 mb & 0.22 mb \\ \hline
& with b-exchange & 6 mb & 14 mb & 1.5 mb & 4.2 mb \\[-1ex]
\raisebox{1.5ex}{$NB_{c}^{-}\rightarrow \overline{D}\Lambda _{b}$} & without b-exchange & 4.5 mb & 10.5 mb & 0.6 mb & 1.5 mb \\ \hline
& with b-exchange & 0.6 mb & 2 mb & 0.7 mb & 2.1 mb \\[-1ex]
\raisebox{1.5ex}{$NB_{c}^{-}\rightarrow \overline{D}^{*}\Lambda _{b}$} & without b-exchange & 0.6 mb & 2 mb & 0.7 mb & 2.3 mb \\ \hline
\hline

\end{tabular}%
\caption{The peak values of the cross sections of the four processes with and without b-flavor exchange, using coupling values of set 1 and 2.}%
\label{table2}%
\end{table}%

\section{Effect of anomalous parity interaction}

The diagrams of the Fig. 1 are produced using PPV, BBP and BBV couplings
defined in Eqs. \ref{2}. However, if the PVV coupling of $B_{c}$ meson due
to anomalous parity interaction is also included then two additional
diagrams shown in Fig. 4 are introduced for the processes (ii) and (iv)
respectively. The effective Lagrangian density defining the anomalous
interaction of mesons\ is discussed in \cite{oh2001}. Here, we report the
relevant interaction term of the Lagrangian density as following.%
\begin{equation}
\mathcal{L}_{B_{c}B^{\ast }D^{\ast }}=g_{B_{c}B^{\ast }D^{\ast }}\varepsilon
_{\alpha \beta \mu \nu }\left[ \left( \partial ^{\mu }D^{\ast \nu }\right)
\left( \partial ^{\alpha }B^{\ast \beta }\right) B_{c}^{-}+B_{c}^{+}\left(
\partial ^{\alpha }\overline{B}^{\ast \beta }\right) \left( \partial ^{\mu }%
\overline{D}^{\ast \nu }\right) \right]
\end{equation}

\begin{figure}[!h]
\begin{center}
\includegraphics[angle=0,width=0.50\textwidth]{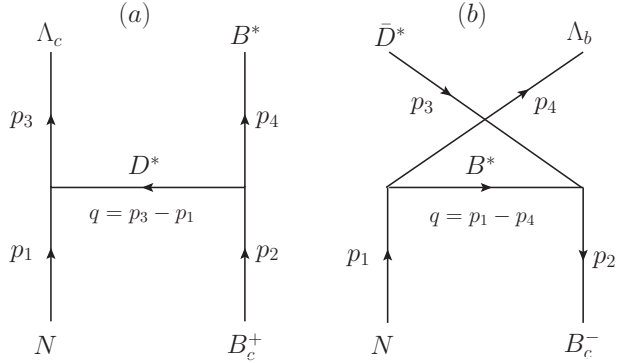}
\end{center}
\caption{Additional Feynman diagrams for $B_{c}$ absorption processes:\ $%
(2)~NB_{c}^{+}\rightarrow\Lambda _{c}B^{\ast }$ and $(4)~NB_{c}^{-}%
\rightarrow \overline{D}^{\ast }\Lambda _{b},$\ due to anomalous parity
interaction.}
\label{fig4}
\end{figure}

The coupling constant $g_{B_{c}B^{\ast }D^{\ast }}$, which has the dimension
of GeV$^{-1}$, can be approximated by $g_{B_{c}B^{\ast }D}/\overline{M}_{D}$
in heavy quark mass limit \cite{hquark}. Where, $\overline{M}_{D}$ is the
average mass of $D$ and $D^{\ast }$ mesons. The scattering amplitudes of the
diagrams of Fig. 4 are given by,%
\begin{eqnarray}
M_{2c} &=&g_{D^{\ast }N\Lambda _{c}}g_{B_{c}B^{\ast }D^{\ast }}\varepsilon
^{\alpha \mu \beta \nu }(p_{4})_{\alpha }(p_{3}-p_{1})_{\beta }\frac{-i}{%
t-m_{D^{\ast }}^{2}}\left( g_{\nu \lambda }-\frac{(p_{3}-p_{1})_{\nu
}(p_{3}-p_{1})_{\lambda }}{m_{D^{\ast }}^{2}}\right)  \notag \\
&&\times \overline{u}_{\Lambda _{c}}(p_{3})\gamma ^{\lambda
}u_{N}(p_{1})\varepsilon _{B^{\ast }}^{\mu }(p_{4}), \\
M_{4c} &=&-g_{B^{\ast }N\Lambda _{b}}g_{B_{c}B^{\ast }D^{\ast }}\varepsilon
^{\alpha \nu \beta \mu }(p_{3})_{\beta }(p_{1}-p_{4})_{\alpha }\frac{-i}{%
u-m_{B^{\ast }}^{2}}\left( g_{\nu \lambda }-\frac{(p_{1}-p_{4})_{\nu
}(p_{1}-p_{4})_{\lambda }}{m_{B^{\ast }}^{2}}\right)  \notag \\
&&\times \overline{u}_{\Lambda _{b}}(p_{4})\gamma ^{\lambda
}u_{N}(p_{1})\varepsilon _{D^{\ast }}^{\mu }(p_{3}),
\end{eqnarray}

Shown in Fig. 5 are the cross sections of the processes (ii) and (iv) with
anomalous interaction for the both coupling sets. The results show that the
cross section of the process (ii) is increased by $\sim 0.2$ mb and $\sim 0.5
$ mb for the coupling sets 1 and 2 respectively, in most part of the energy
range. Fig. 5 also shows that the effect of anomalous interaction on cross
section of the process (iv) is negligible for the both coupling sets.
Although the anomalous interaction significantly increases the cross section
of the process (ii), but this effect is marginal on the total cross section
(i.e., the sum of four processes) due to relatively small value of the cross
section of the process (ii).

\begin{figure}[!h]
\begin{center}
\includegraphics[angle=0,width=0.45\textwidth]{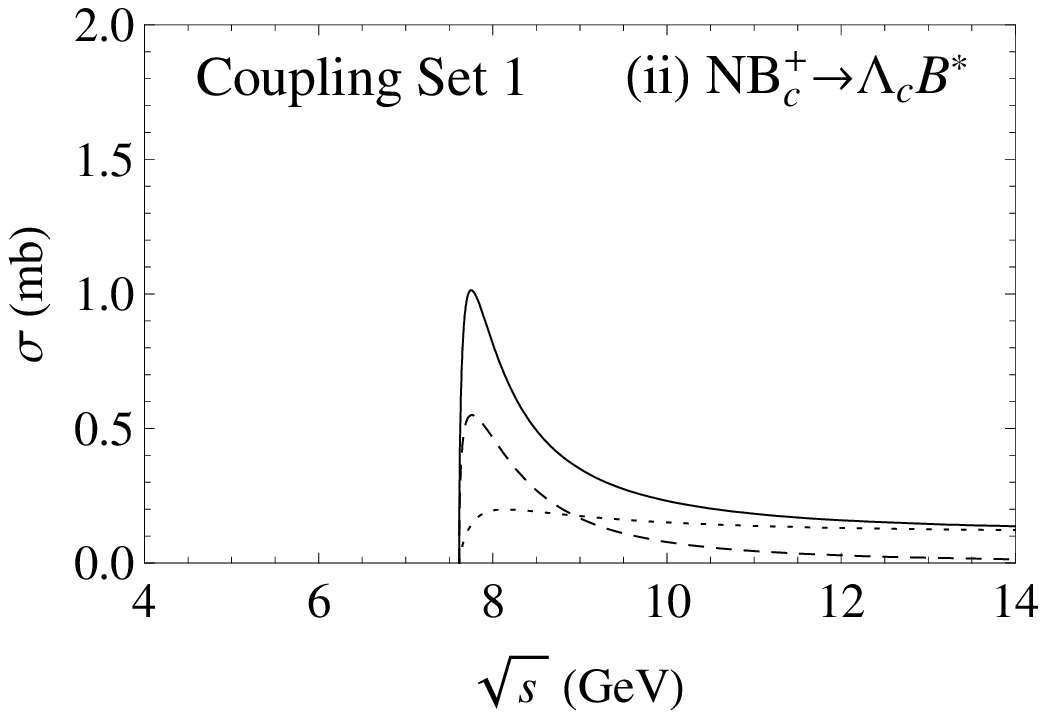} \label{fig5a} %
\includegraphics[angle=0,width=0.45\textwidth]{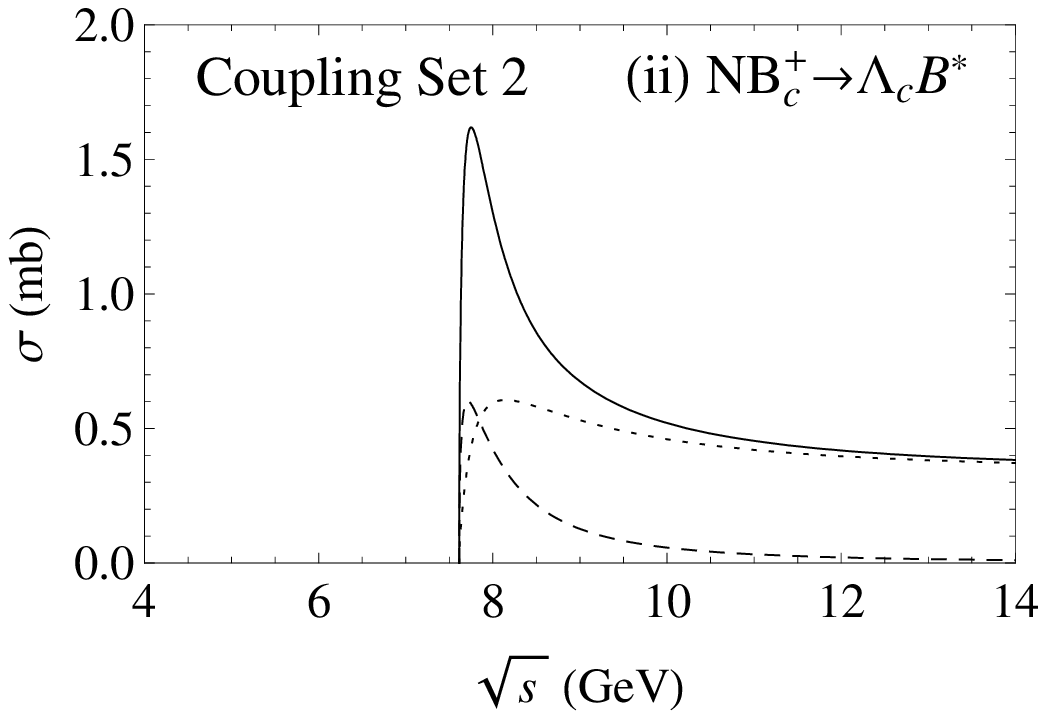} \label{fig5b}
\includegraphics[angle=0,width=0.45\textwidth]{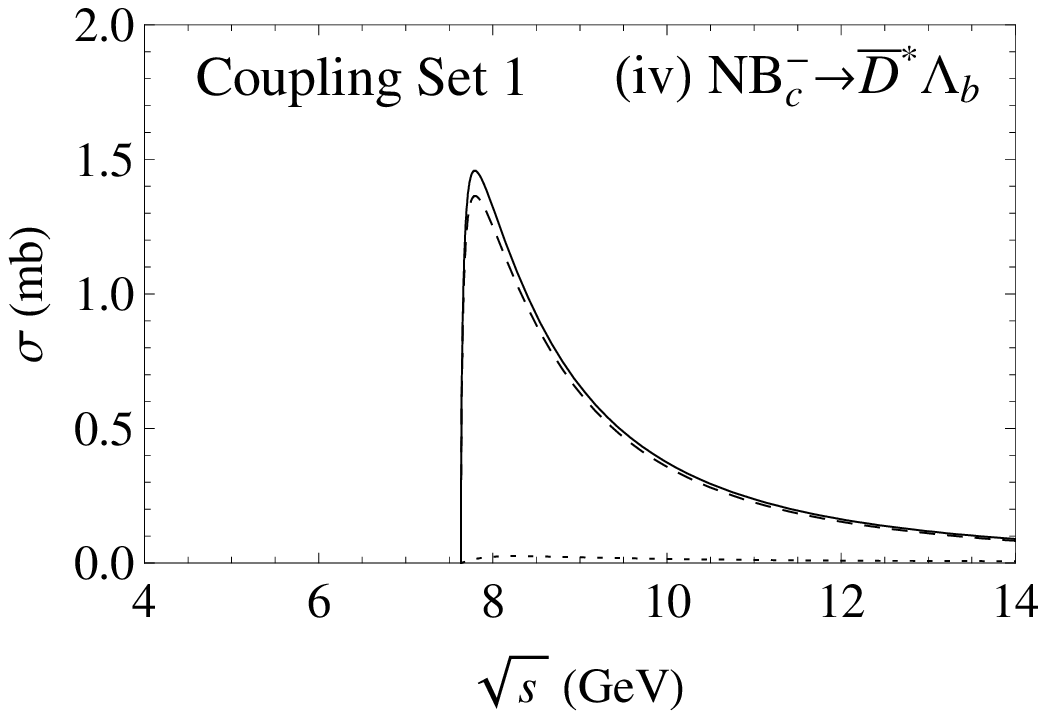} \label{fig5c} %
\includegraphics[angle=0,width=0.45\textwidth]{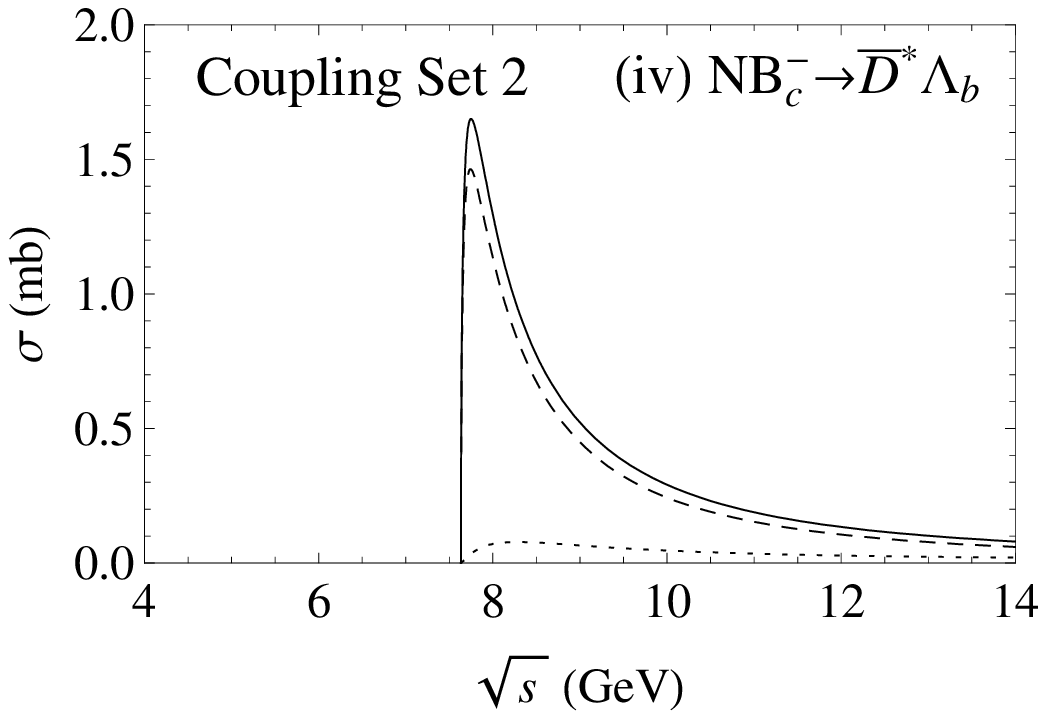} \label{fig5d}

\end{center}
\caption{$B_{c}$ absorption cross sections of the processes (ii) and (iv) using the values of the couplings given in set 1 and set 2. Solid and dashed curves represent the cross sections with and without anomalous diagrams respectively, and dotted curves represent the contribution from anomalous diagrams alone, i.e., without including the contribution from the interference terms. Cutoff parameter $\Lambda$ is taken $1.5$ GeV.}

\label{fig5}
\end{figure}

\section{Concluding Remarks}

\noindent In this paper, we have calculated $B_{c}$ absorption cross
sections by nucleons in meson-baryon exchange model using hadronic
Lagrangian based on SU(4)/SU(5) flavor symmetries. This approach has already
been used for calculating absorption cross sections of $J/\psi $ and $%
\Upsilon $ mesons by pions, $\rho $ mesons and nucleons. In order to
calculate $B_{c}$ absorption cross sections, we use PPV, BBP and BBV
couplings given in Eq. \ref{2}. Related coupling constants $g_{DN\Lambda
_{c}}$ and $g_{D^{\ast }N\Lambda _{c}}$ can either be fixed by SU(4)/SU(5)
flavor symmetries or empirically using QCD-sum rules. We have calculated
absorption cross sections using both set of values for comparison. Whereas,
for the coupling constants $g_{BN\Lambda _{b}}$, $g_{B^{\ast }N\Lambda _{b}}$
and $g_{B_{c}\Lambda _{c}\Lambda _{b}}$ no empirical values are available,
so we use SU(5) symmetry relations. These estimates are less reliable as the
SU(5) flavor symmetry is broken due to large variation in the quark masses.
It is noted that for the processes (i) and (ii) only the b-flavor exchange
diagrams depend upon these three couplings. Thus the effect of these
couplings on the cross sections of the first two processes is less
significant when the contribution of b-flavor exchange diagram is small or
negligible as in the case of process (i) for coupling values of the set 2.
However, for the processes (iii) and (iv), the amplitudes of both c and
b-flavor exchange diagrams depends upon these couplings. Thus, any change in
the values of these couplings could significantly change the cross sections
of these processes irrespective of the relative contribution of the two
diagrams. We conclude that a more rigorous study on these couplings could
further improve our results. The anomalous parity interaction is found to be
significant only for the process (ii). The effect, however, itself marginal
in total value of the cross section due to lesser contribution from the
process (ii). 

\appendix

\section{Appendix}

\noindent In SU(5) quark model, $J^{P}=\frac{1}{2}^{+}$ baryons (anibaryons)
are $40$ ($\overline{40} $)-plets of 1100 (0011) representation and mesons
are 24-plets of 1001 representation of SU(5) group, whereas SU(5) invariant
Lagrangian defining BBP or BBV couplings must be a singlet. Since $40\otimes
24=450\oplus 210\oplus 175\oplus 40\oplus 40\oplus 35\oplus 10$, \ there are
two possible BBP and BBV couplings as in case of SU(3) and SU(4). These
SU(5) invariant couplings are expressed in terms of irreducible tensors $%
P_{\alpha }^{\beta } $, $V_{\alpha }^{\beta }$ and $\phi _{\mu \nu \lambda }$
in Eqs. \ref{4} and \ref{5}. The tensor $\phi _{\mu \nu \lambda }$, which
define $J^{P}=\frac{1}{2}^{+}$ baryons, satisfies the conditions $\phi _{\mu
\nu \lambda }+\phi _{\lambda \mu \nu }+\phi _{\nu \lambda \mu }=0$ and $\phi
_{\mu \nu \lambda }=\phi _{\nu \mu \lambda }$. The relations defining $J^{P}=%
\frac{1}{2}^{+}$ baryons in terms of the elements of $\phi _{\mu \nu \lambda
}$ for $u,d,s\,\ $and $c$ quarks are given in ref. \cite{Okubo1975}. Here,
we present the relations defining the baryons with bottom quark(s).
\begin{gather}
\Sigma _{b}^{+}=\phi _{115},\ \ \ \Sigma _{b}^{0}=\sqrt{2}\phi _{125},\ \ \
\Sigma _{b}^{-}=\phi _{225},  \notag \\
\Xi _{b}^{0}=\sqrt{2}\phi _{135},\ \ \ \Xi _{b}^{-}=\sqrt{2}\phi _{235},
\notag \\
\ \ \Xi _{b}^{\prime 0}=\sqrt{\frac{2}{3}}\left( \phi _{513}-\phi
_{531}\right) ,\ \ \ \Xi _{b}^{\prime -}=\sqrt{\frac{2}{3}}\left( \phi
_{523}-\phi _{532}\right) ,  \notag \\
\Lambda _{b}^{0}=\sqrt{\frac{2}{3}}\left( \phi _{521}-\phi _{512}\right) ,\
\ \ \Omega _{b}^{-}=\phi _{335},\ \ \ \Omega _{ccb}^{+}=\phi _{445},  \notag
\\
\Xi _{bb}^{0}=\phi _{155},\ \ \ \Xi _{b}^{-}=\phi _{552},  \notag \\
\Omega _{bb}^{-}=\phi _{553},\ \ \ \Omega _{bbc}^{o}=\phi _{554},  \notag \\
\Xi _{cb}^{+}=\sqrt{2}\phi _{145},\ \ \ \Xi _{cb}^{0}=\sqrt{2}\phi _{245},
\notag \\
\Xi _{cb}^{\prime +}=\sqrt{\frac{2}{3}}\left( \phi _{514}-\phi _{541}\right)
,\ \ \ \Xi _{cb}^{\prime 0}=\sqrt{\frac{2}{3}}\left( \phi _{524}-\phi
_{542}\right) ,  \notag \\
\Omega _{cb}^{0}=\phi _{345},\ \ \ \Omega _{cb}^{\prime 0}=\sqrt{\frac{2}{3}}%
\left( \phi _{543}-\phi _{534}\right) ,
\end{gather}

\end{document}